\documentclass[]{rsos}

\usepackage{float}
\usepackage{amsmath}    
\usepackage{amssymb}
\usepackage{multirow}
\usepackage{units}
\usepackage{ulem}
\usepackage[table]{xcolor}
\usepackage[final]{pdfpages}

\definecolor{orange}{RGB}{253,245,230}

\newcommand{\new}[1]{{\color{black}}}

\begin{document}

\title{Evolution in the Debian GNU/Linux software network: analogies and differences with gene regulatory networks}

\author{
Pablo Villegas$^{1,2}$, Miguel A. Mu\~noz$^{1}$ and Juan A. Bonachela$^{3,4}$}

\address{$^{1}$Departamento de Electromagnetismo y F{\'i}sica de la Materia and Instituto Carlos I de F{\'i}sica Te{\'o}rica y Computacional. Universidad de Granada, E-18071 Granada, Spain\\
$^{2}$Istituto dei Sistemi Complessi, CNR, via dei Taurini 19, 00185 Rome, Italy\\
$^{3}$Marine Population Modeling Group, Dept. of Mathematics and Statistics, University of Strathclyde, Glasgow, G1 1XH, Scotland, UK\\
$^{4}$Department of Ecology, Evolution, and Natural Resources, Rutgers University, New Brunswick, 08901 NJ, USA}

\subject{}

\keywords{evolving networks, network resilience, information transmission}

\corres{Juan A. Bonachela\\
\email{juan.bonachela@rutgers.edu}}

\begin{abstract}
Biological networks exhibit intricate architectures deemed to be crucial for their functionality. In particular, gene regulatory networks, which play a key role in information processing in the cell, display non-trivial architectural features such as scale-free degree distributions, high modularity, and low average distance between connected genes. Such networks result from complex evolutionary and adaptive processes difficult to track down empirically. On the other hand, there exists detailed information on the developmental (or evolutionary) stages of open-software networks that result from self-organized growth across versions. Here, we study the evolution of the Debian GNU/Linux software network, focusing on the changes of key structural and statistical features over time. Our results show that evolution has led to a network structure in which the out-degree distribution is scale-free and the in-degree distribution is a stretched exponential. In addition, while modularity, directionality of information flow, and average distance between elements grew, vulnerability decreased over time. These features resemble closely those currently shown by gene regulatory networks, suggesting the existence of common adaptive pathways for the architectural design of information-processing networks. Differences in other hierarchical aspects point to system-specific solutions to similar evolutionary challenges.
\end{abstract}


%


\maketitle

\section{Introduction}
Understanding the collective properties stemming from the interactions of a large number of units such as genes, proteins, or metabolites is of paramount importance  in biology \cite{benner2005,purnick2009,khalil2010_ch6}. Theoretical work focusing on the changes over time of self-organizing networks can provide key information about these natural systems. Particularly, network theory provides us with a highly-insightful systems-level perspective to extremely complicated biological problems, which has helped advance knowledge in fields such as neuroscience \cite{Hagmann2008}, ecology \cite{Sole2001}, and epidemiology \cite{Pastor2001}, to name a few \cite{newman2003_review}. The study of information processing in living systems has greatly benefited from this network perspective, complementing parallel endeavors for the analysis of single pathways, and providing a much richer understanding of collective phenomena emerging from a large number of basic inter-related units \cite{Kitano,Alon}.

More specifically, analyses of gene-regulatory, protein-protein, and metabolic networks have led to dramatic advances in systems biology \cite{Jeong2000,barabasi2004_Rev}. Indeed, an important step forward in the understanding of cell regulatory mechanisms was the discovery of the scale-free degree distribution of such networks \cite{barabasi2004_Rev}. Most networks within the cell show other non-trivial structural features such as a small-world property (arising from an small average path length, or distance between any two elements), a highly modular structure, and an extraordinary responsiveness to
environmental cues, as well as dynamical and structural robustness \cite{barabasi2004_Rev}.  In particular, gene regulatory networks (GRNs) --which display mutual regulatory interactions between genes-- have been identified as directed networks that exhibit a scale-free distribution for the number of regulated genes (called ``out degree'' due to the direction of information flow) but an exponential distribution for the number of controlling genes (``in degree'').
In addition, cellular states can be identified as attractors of the dynamics of genetic regulatory systems, which allows the latter to be modeled --at least in first approximation-- as random Boolean networks \cite{kauffman69_ch6,kauffman1993_ch6,gros2011_ch6,villegas2016}. In such a simplified approach, mutual regulatory interactions, described as direct links between genes/nodes, involve arbitrary random Boolean functions whose inputs are the on/off states encoding the expression level of other genes. This type of binary setup has shed light on important conceptual problems such as the emergence of diverse phenotypes from a unique genetic network, the existence of transitions among them (e.g. cell differentiation and reprogramming), and the emergence of cycles in cell states.  In this regard, two examples are the predicted expression patterns of the fly \textit{Drosophila melanogaster} \cite{droso_ch6} and the yeast cell cycle \cite{yeast_ch6}. In either case, it seems clear that gene-regulatory networks involve some type of information transmission (or flow) encoded in mutual regulatory interactions, determining the cellular response to different stimuli or environmental conditions \cite{tkavcik2008,barabasi2004_Rev}. The analysis and study of interacting systems with similar information flow can help understand the particular structure and emergent properties of genetic systems.

One particular aspect of GRNs that remains elusive is how such properties and functionality have emerged through evolution. The main reason for this knowledge gap is that the information currently available provides a limited picture of the evolutionary path followed by biological networks. Artificial, self-organized networks, on the other hand, can offer an unrivaled level of detail regarding their different ``evolutionary'' stages. In this respect, self-organized software networks have been shown to constitute an excellent model for the study of the evolution of biological networks \cite{Fortuna2007,Fortuna2011,valverde2005_ch6,pang2013_ch6,keil2018,valverde2015}.

Software networks are composed by packages, acting as nodes and conforming the basic unit of software. These are inter-related due to the need for a package to reuse code of other packages in order to work properly (the so-called dependencies, i.e. package $i$ needs other packages to be functional). Similarly to GRNs, the storage and transmission of information through this network enables the proper functioning of the operating system as a whole. Thus, it is pertinent to ask: what can the characterization of the evolutionary history of a software network reveal about the evolution and main architectural features of gene regulatory networks?

To shed light onto this question, we focus here on the Debian GNU/Linux operating system (Debian, hereon), for which publicly available historical data exist.  First, we characterize the evolving structure of the network of dependencies between software packages in the different Debian distributions released up to June $2019$. Then, we explore the emergent properties of such networks and their role in the functionality of the system. With this information, we finally scrutinize the similarities and differences between the structure and emergent properties of software networks and GRNs.

\section{Debian networks} 

Debian is an open-source operating system that, in the last $25$ years, has changed in a self-organized way through the collective action of a myriad of developers. The history of Debian shows many small intermediate steps but only $14$ important stable releases (evolutionary steps) that have progressively altered its networked structure. This evolution has resulted in a sustained growth generating an actual network of interactions between thousands of packages, starting from a very small initial release in its first version (see Fig.\ref{fig:Sketch}). Such packages must translate the information coming from other packages in order to satisfy the so-called dependencies, i.e. pieces of software required for the package to work. Thus, we can represent each version of the operating system using a different network for each of the different evolutionary stages. Nodes represent packages and links join every node with the packages it depends upon, thus forming a directed network of dependencies \cite{Fortuna2011}. Note that the specific interpretation of the flow of information determines the direction of the links. Flow in terms of dependencies imposes that directional links start on a focal package and point to its dependencies (e.g. \cite{Fortuna2011}). Flow in terms of information transmission, like in GRNs, is typically represented with opposite directional links: because {\it i} depends on {\it j}, information flows from {\it j} to {\it i}. Here, we use the latter convention to facilitate comparison between the Debian networks and GRNs. 

We have focused on the relationships between the binary \textit{x}86 packages included in Debian, for its first $14$ (stable) distributions Figure \ref{fig:Sketch}, left). Figure \ref{fig:Sketch} (right panel) shows one specific example of package inter-dependence extracted from the very first distribution. In addition to dependencies reflecting requirements, additional relationships between packages reflect incompatibilities among them (i.e. ``conflicts'') \cite{Hertzog2014}. In most cases, conflicts are used to avoid duplication in virtual facilities of the system, as explained in the Debian Policy Manual \cite{DPolicy}. Most conflicts occur between potential candidates to fulfill a requirement or function, grouped under the term ``virtual packages''. Such virtual facilities are supplied by particular choices among many possibilities (e.g. both \textit{firefox} and \textit{konqueror} provide the same service, \textit{www-browser}). Here, we resolve conflicts by averaging over the ensemble of potential networks resulting from random choices for each ``virtual package'' (see Methods), which allows us to focus only on the network of mutual dependencies.

\begin{center} \begin{figure}[hbtp] 
\begin{centering} \includegraphics[width=0.95\columnwidth]{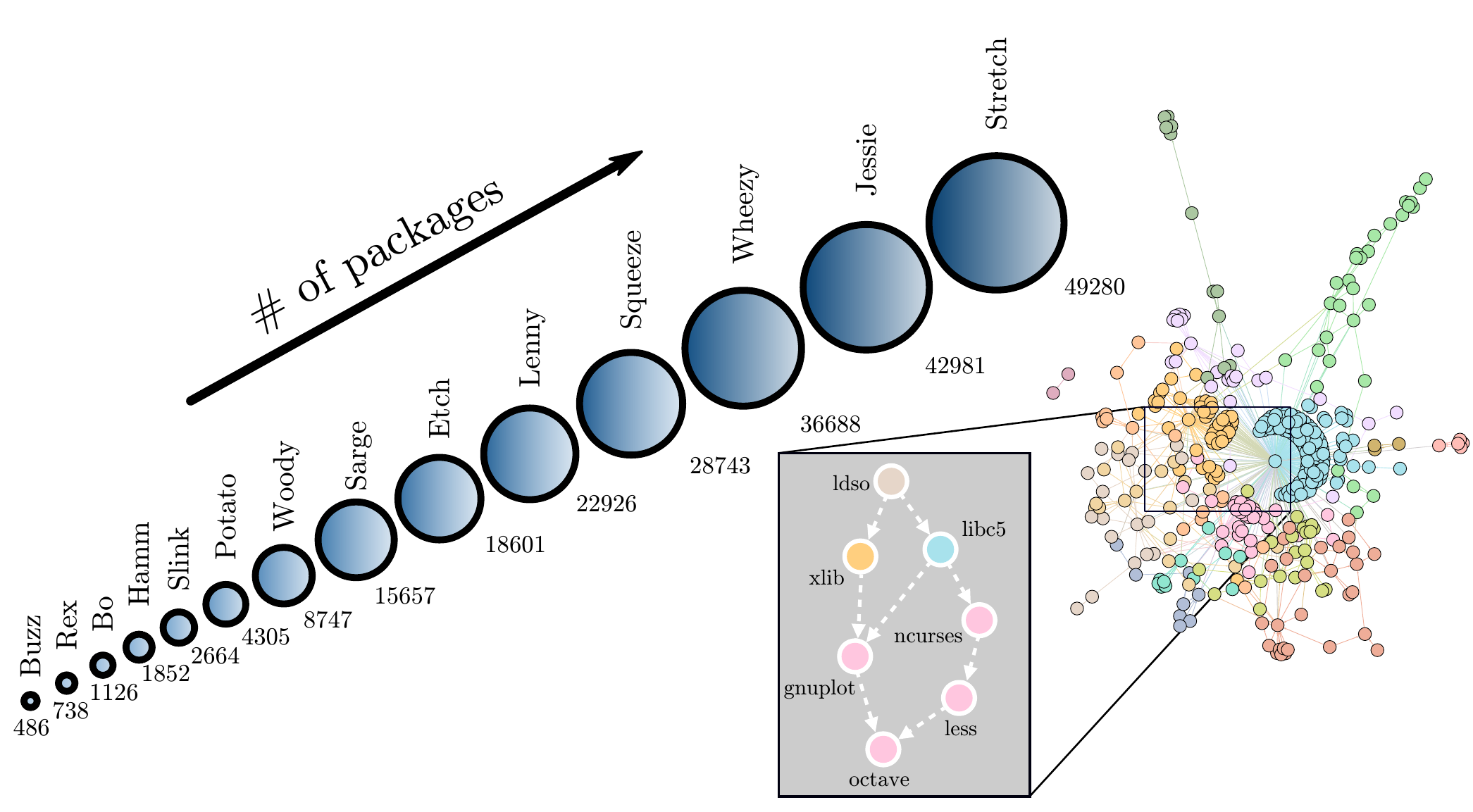} \par\end{centering} \caption{Left: Sketch showing the official code-name and the total number of packages of all stable Debian releases (releases named after characters from the movie {\it Toy Story}\textcopyright); circle sizes are proportional to the number of packages in each distribution. Right: Sample of the dependencies network for the first release, where different colors indicate package moduli in the network (which we do not specify here); inset: subgraph showing specific dependencies between some well-known packages from this distribution.\label{fig:Sketch}} \end{figure} \par\end{center}%

\section{Results} 

Since the initial Buzz distribution, which included $486$ packages, the total number of packages conforming each Debian stable version has grown exponentially (see Fig.\ref{fig:Sketch} and Supplementary Information 1, SI-1). Here, we scrutinize this growth across releases (evolutionary steps), and analyze how it translated into the emergent structural properties of the associated Debian software networks.

\subsection*{Degree distributions}

\begin{figure}[hbtp]
\centering \includegraphics[width=0.95\columnwidth]{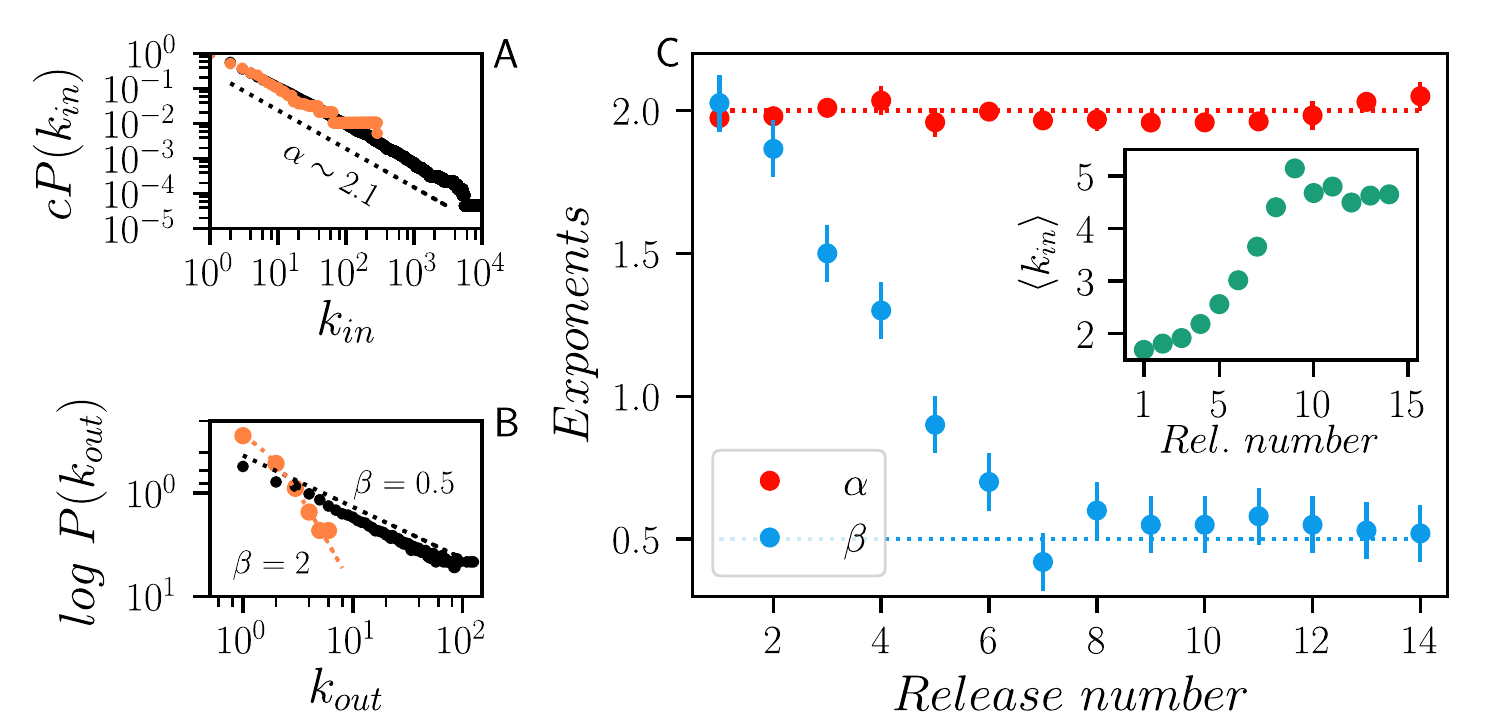}
\caption{A) Cumulative degree distribution for outgoing dependencies for the first (orange) and last (black) releases in log-log scale. B) Logarithm of the degree distribution for in-degree for the same two releases, showing a half-normal distribution decay ($\beta=2$) for the first version and a stretched exponential distribution ($\beta=0.5$) for the last one. C) Exponents from the out-degree distribution of dependencies power law (blue) and the in-degree distribution stretched exponential (red) for the first release to the most recent one. Inset: Mean in-degree for all distributions. 
\label{FigExp}}
\end{figure}

To characterize the changes in the structural properties of Debian over time, we first analyzed the in- and out-degree distributions (averaged over realizations of the network obtained after resolving conflicts). The former describes the probability for a package to depend on $k_{in}$ packages, whereas the latter describes the probability for a package to be a requirement for $k_{out}$ packages. As shown in Fig.\ref{FigExp}A and Fig.\ref{FigExp}B (respectively), the cumulative out-degree distribution exhibits a power law tail, $P(k_{out})\sim k_{out}^{-\alpha+1}$ with characteristic exponent $\alpha$, and the in-degree distribution follows a stretched exponential, $P(k_{in})\sim \exp(-(\frac{k_{in}}{\tau})^{\beta})$ with characteristic exponent $\beta$ (see Methods, and SI-2 for fits and estimations of their likelihood). As shown in Fig.\ref{FigExp}C the power law for the out-degree distribution shows an exponent very close to $\alpha=2$ for all releases (red points). The exponent of the stretched exponential for the in-degree distributions (blue points) decays from $\beta=2$ in the early releases (normal distribution), to an approximately-stationary value $\beta\approx0.5$ after the $8$th release. The inset of Fig.\ref{FigExp}C shows that the mean in-degree connectivity, $\left\langle k_{in}\right\rangle$, grows from $1.5$ to an apparently-stationary value $\sim4.75$, reached after the 8th release.

\subsection*{Emergence of a non-trivial modular structure}

\begin{figure}[hbtp]
\centering \includegraphics[width=0.95\columnwidth]{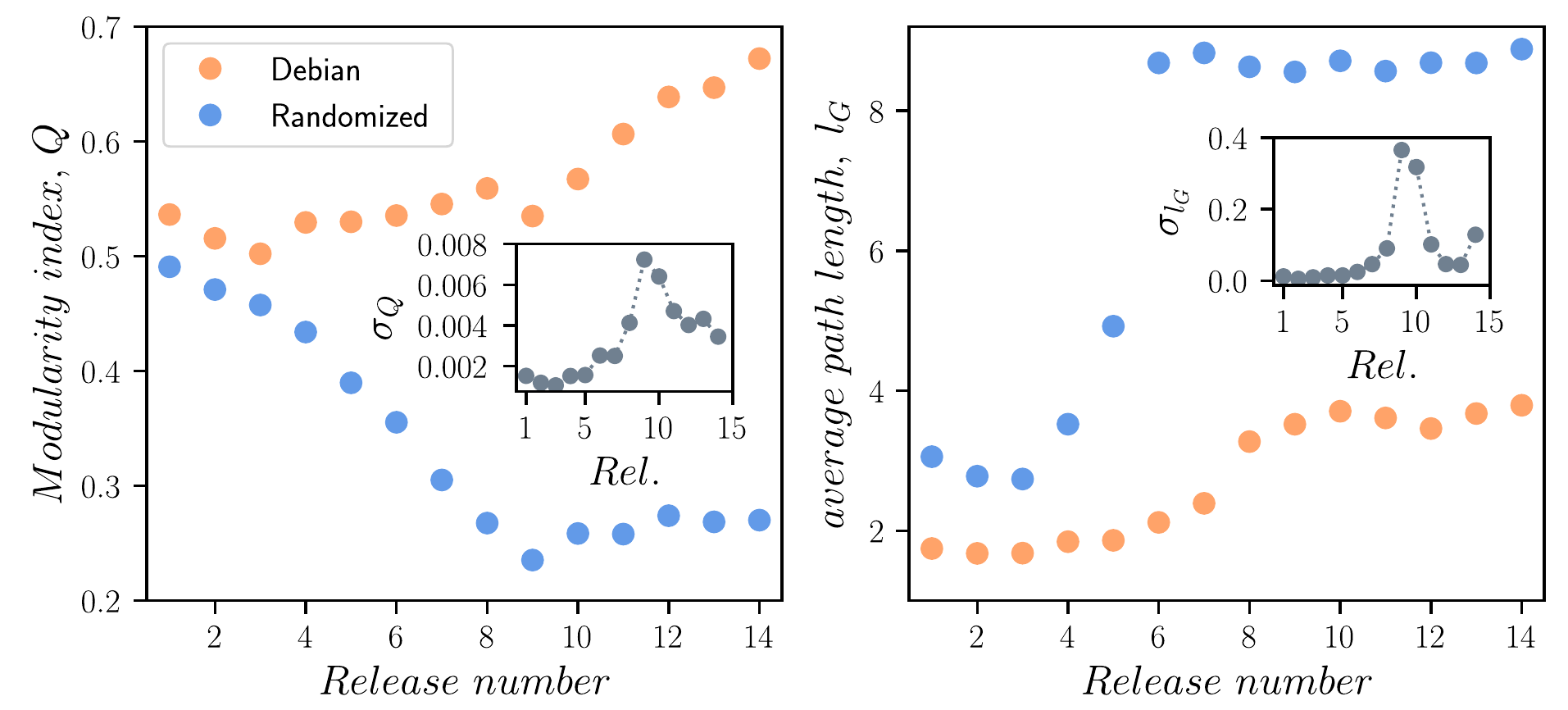}
\caption{(Left) Modularity index ($Q$) for the different Debian releases (orange points) and the ensemble of random (``swapped'') networks (blue points). (Right) Average path length ($l_G$) for Debian networks (orange points) and the ``swapped'' networks (blue points). In both panels, insets show the standard deviation for the ``swapped'' networks $Q$ and $l_G$ are averaged over $10^4$ realizations.
\label{FigMod}}
\end{figure}

We also quantified computationally how modular is the structure of each release's network, for which we used the Newmans's modularity index\cite{newman2003_review} ($Q$, see Methods). As Fig. \ref{FigMod} (left) shows, modularity grows continuously for the last six versions (i.e. after the $8$th release), from $0.5$ to $\sim0.7$ (orange points). We further explored the possibility for such modularity to be an artifact of the degree sequence of the nodes. To this end, and although the Newman's modularity index already discounts the random expectation for random networks, we constructed null/random structure networks by randomizing the original networks exchanging links between random nodes under certain rules. Specifically, for each original (conflict-resolved) Debian network ($D$), we constructed a random copy ($D'$) in which a swapping process maintained both the incoming and the outgoing connectivity of each single node unaltered (see Methods). As illustrated by the blue points in Fig. \ref{FigMod} (left), the randomized networks show a sustained decrease of their modularity index until the $8$th distribution. The standard deviation of the modularity index for the ``swapped'' networks shows a pronounced peak around such version as well (see inset). Moreover, the randomization allowed us to calculate a Z-score for $Q$ to estimate the difference in modularity between the Debian networks and their randomized counterparts (normalized by the available diversity, i.e. standard deviation, of the ensemble of possible networks, see Methods). This Z-score shows a steep increase over time, i.e. the modularity of Debian becomes progressively larger than that of its random-network counterparts over time (see Fig. S4).

We also measured how the average path length ($l_G$) changed across the different Debian releases. Figure \ref{FigMod} (right) shows a small increase in $l_G$ for the original networks (orange points) constrained in $\left(1.75-3.8\right)$. This growth slows down after version $8$. For the ``swapped'' networks (blue points), $l_G$ shows a steep increase that jumps to a considerably higher value ($\sim8.75$) around the $6$th version, and remains at that level afterwards. This gap approximately coincides with a pronounced peak for its standard deviation in the ensemble of randomized networks (see inset). The Z-score for the average path length grows monotonically, showing two growth regimes before and after the $8$th release (see Fig. S4). As the size of the network increases, the distance that information needs to travel between any two packages in Debian becomes larger on average, but grows much more slowly than that for the randomized counterparts. Moreover, Debian grew accordingly to the expectation for a typical small-world network (i.e. proportionally to the logarithm of the number of elements, see Fig. S5), with different trends before and after the 8th release.

\vspace{0.5cm}
\subsection*{Evolving hierarchy}

To further analyze the evolution of the network structure, we explored how diverse aspects of the hierarchy of the Debian networks changed with time. Due to the multi-faceted character of hierarchy, there is a lack of a single definition or observable to measure \cite{Corominas}, and hence we monitored three different indicators: {\it a)} the level of ``stratification'' or re-use of software in the network; {\it b)} the directionality of the information flow; and {\it c)} the level of reciprocal dependence between packages (see Methods).

\begin{figure}[hbtp]
\centering \includegraphics[width=1.0\columnwidth]{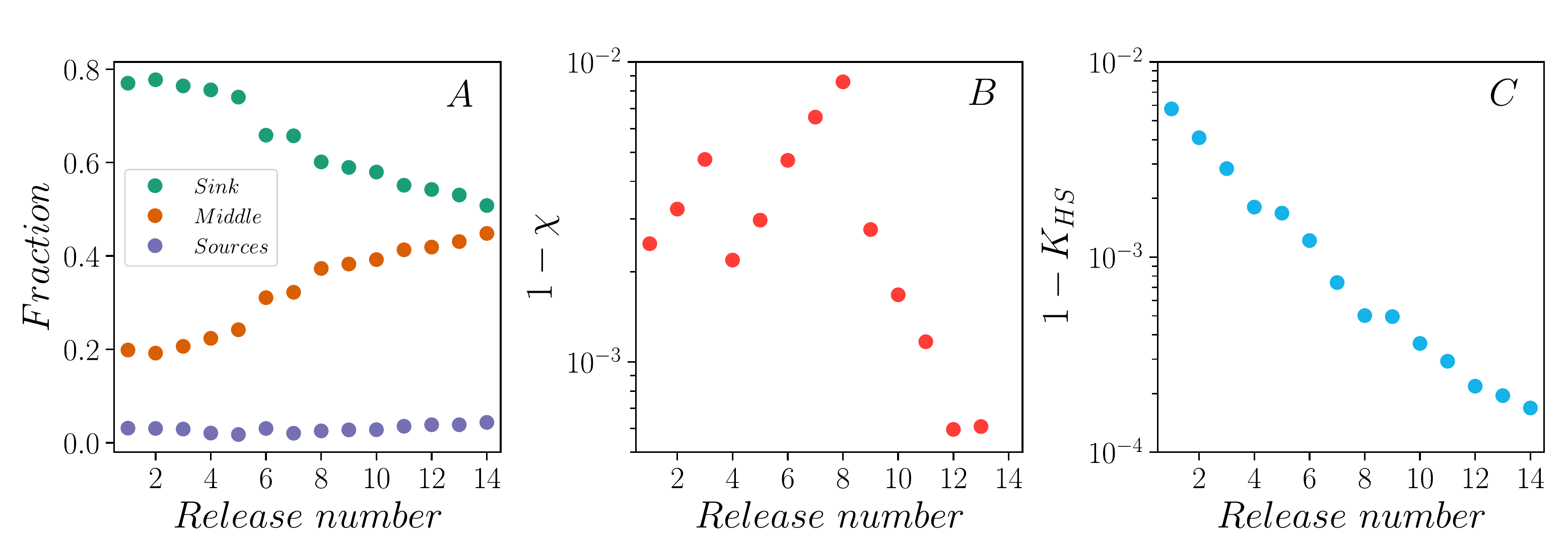}
\caption{(A) Evolution of the three main hierarchical categories defined in the network: ``sinks'', ``pass-through nodes'' and ``sources''. Observe that the ``sources'' of information, i.e. nodes that not depend on any node, constitute a small fraction of the network. However, ``sinks'', situated in the lowest point of the hierarchy, are disappearing over time. (B) Flow index (represened as $1-\chi$ to allow for the semi-log plot), showing an initial discontinous decrease followed by a monotonic increase after the $8th$ version. (C) Reciprocal of the Krackhardt hierarchy score, $1-K_{HS}$, which decreases monotonically for all the Debian releases. Altogether points to a dynamic change of the various aspects of hierarchy for the network.\label{FigHi}}
\end{figure}

To measure the stratification/re-use of software in the network, we introduced three categories for packages \cite{Yan2010}: i) packages with no dependencies (``sources'' or information containers); ii) packages that depend on other packages and, at the same time, some packages depend on them (``pass-through'' nodes), and iii) packages that no other package depends upon (``sinks''). In this way, each network is decomposed into three distinct hierarchical levels or strata. We monitored the fraction of packages within each of these hierarchical categories across the different releases.As shown in Figure \ref{FigHi}A, ``source'' packages only represent a small fraction of any given network, whereas the proportion of ``sinks'' decreases progressively from a remarkable $0.8$ value observed for the first distributions. At the same time, the number of ``pass-through'' nodes increases, indicating that most sinks moved to this category (i.e. packages tend to be re-used).

To quantify the directionality of information flow, we measured the so-called flow index, $\chi$ \cite{VirIndex}. In a nutshell, $\chi$ quantifies the fraction of links pointing from lower to higher hierarchical levels, i.e. fraction of links that are aligned with an inherent directionality. A value of $\chi$ close to $1$ indicates that there is an overall directional flow such that a large fraction of links point from higher to lower hierarchical levels. As Fig. \ref{FigHi}B shows, Debian overall exhibits high directionality (i.e. low $1-\chi$ values). Although, for earlier releases, the fraction of links pointing in the direction of the flow decreases discontinuously, the trend inverts after the 8th release, and $1-\chi$ shows a steep decrease (i.e. increase of directionality) over the following versions of Debian.

To quantify the reciprocal dependence between packages, we measured a classic hierarchy index, the Krackhardt hierarchy score, $K_{HS}$ \cite{Krackhardt}. Here, we represented $1-K_{HS}$, which indicates the fraction of pairs that show mutual or reciprocal dependence (i.e. package {\it i} can be reached from package {\it j} and the other way around). Fig. \ref{FigHi}C shows a monotonic decrease for $1-K_{HS}$, representing a loss of symmetrically linked pairs in the network across releases.

In summary, the three measures altogether show that the hierarchy of the Debian networks has changed over time, strengthening the overall directionality of the flow and keeping a low reciprocal dependence, while increasing the re-use of software.

\vspace{0.5cm}
\subsection*{Evolution of robustness}
\begin{figure}[hbtp]
\centering \includegraphics[width=0.6\columnwidth]{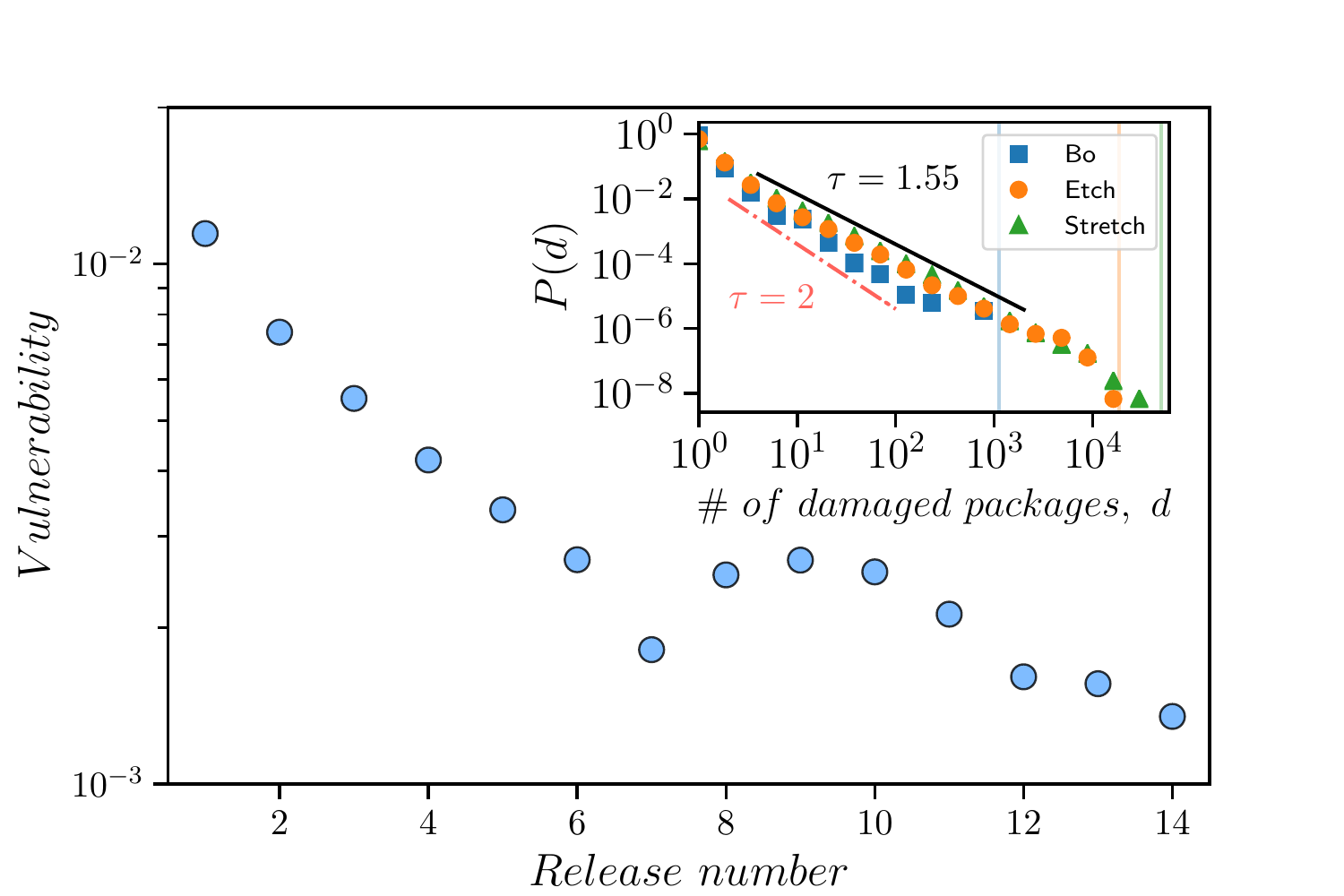}
\caption{Vulnerability index for the different Debian distributions \new{obtained after $10^{4}$ ``damage'' experiments}, showing an overall decrease of vulnerability. Inset: Scale-free avalanches of damage after an attack to one random node of the network for three representative releases (averaged over possible initial seeds and networks in the ensemble). Lines are guides to the eye for the slopes indicated by the exponents (dot-dashed red line for $\tau=2$, and black lines for $\tau=1.55$). Vertical lines represent the total number of packages in each distribution, which is indicative of upper cutoffs. \new{Error bars are smaller than the symbols}.
  \label{FigAtt}}
\end{figure}

One plausible expectation for the development of any operating system is that its associated network should increase over versions its ability to retain functionality in spite of damages, i.e. it is expected to enhance its robustness. To explore this possibility, we perturbed each Debian network by removing at random one package (i.e. we made a package unusable), and followed the consequent cascade of failures by removing nodes depending upon removed packages, which ultimately affects a certain fraction of the whole network. We called this fraction of affected nodes ``vulnerability''. More precisely, we defined vulnerability as the fraction of affected packages averaged over initial damage ``seeds'' and network ensemble. Figure \ref{FigAtt} shows that the vulnerability index decreases across versions, and exhibits a slow-down for this trend after a peak around the $8$th-$9$th distributions. 

We also analyzed the full probability distribution for the number of packages affected by each cascade of failures (or avalanches size, using the jargon from damage spreading analyses \cite{MAM-rmp}). This avalanche-size distribution shows a power-law tail with an exponent that decreases over time (from an initial $\tau=2$ to $\tau=1.55$ for the last distribution, inset of Figure \ref{FigAtt}).

\vspace{0.5cm}
\subsection*{Network information content}

We also studied the relationship between the total size (measured in megabytes) of the different packages, as a proxy for the total amount of information contained in the Debian network. In particular, we analyzed the evolution of the sum of the sizes of all packages for each potential network as a function of the number of packages (see Figure \ref{FigPack}). As
happens with other properties, the information content grows monotonously across time, but it changes its slope around the $8$th distribution, moving from a linear to a superlinear trend (i.e. the information content of the network grows faster than the number of packages).

\begin{center}
\begin{figure}[hbtp]
\begin{centering}
\includegraphics[width=0.4\paperwidth]{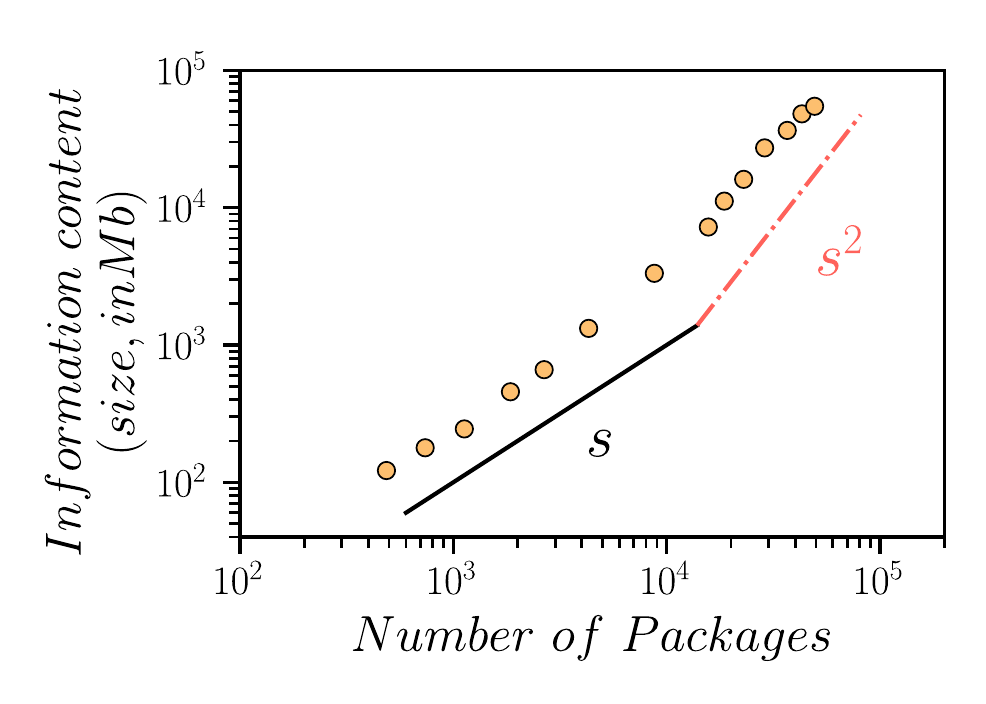}
\par\end{centering}
\caption{Total number of packages  measured as a function of the system size for each Debian release. After the initial, approximately-linear growth, a slowing down can be appreciated as general trend. Lines are guides to the eye showing a linear growth (black) and a quadratic growth (red).
\label{FigPack}}
\end{figure}
\par\end{center}

\subsection*{Results for gene-regulatory networks}

We analyzed gene regulatory networks (GRNs) from publicly available datasets for \textit{E.Coli, M. Tuberculosis, B.Subtilis, P.Aeruginosa} and \textit{S.Cerevisiae} \cite{gama2015_EColi,albergante2014_MTub,arrieta2015_BSubtilis,ma2014_SCerevisiae}. Below, we summarize our main findings, and refer the reader to SI-4 and SI-5 for a comprehensive presentation of all these results.

GRNs are very sparse networks, with mean incoming connectivity around $1.5-3.5$ in most of the cases \cite{villegas2016}. \new{They show a total number of links for a given total number of nodes similar to that of the Debian networks (Fig. \ref{Fig:Comparison}, panel A, were colored nodes of diverse shapes correspond to GRNs).} The distribution for the in-degree --the number of regulatory interactions in which a given gene intervenes-- has been reported to be exponential \cite{barabasi2004_Rev}, whereas the distribution for the out-degree --the number of interactions controlling the expression of a given gene-- is scale free\cite{barabasi2004_Rev}. Our own analyses reveal that the out-degree distribution decays as a power law, $P(k_{out})\sim k_{out}^{-\alpha}$ with characteristic exponent $\alpha \in (1.5-2)$ \cite{Aldana2007_ch6,Jeong2000,buchanan2010_ch6}, and the in-degree distribution follows a stretched exponential, $P(k_{in})\sim \exp(-(\frac{k_{in}}{\tau})^{\beta})$ with characteristic exponent $\beta \in (0.5-1.5)$ (see SI-4). \new{These values are similar to those shown by the Debian networks, given their respective network sizes (panel B).} For the analyzed GRNs, the Newman's modularity index $Q$ is in the interval $\left(0.55-0.8\right)$, and the average path length $l_{G}$ in $\left(2.0-3.5\right)$ \new{(see Fig. \ref{Fig:Comparison}, panels C and D), values that are lower than those of the Debian networks (panels C and D)}. On the other hand, GRNs show a very prominent hierarchical structure: more than $85\%$ of the nodes are ``sinks'', i.e. they do not control other genes\cite{ravasz2002,Yan2010}, there is a very strong directionality of the information flow (i.e. very small $1-\chi$ value), and the fraction of reciprocally-expressed genes, $1-K_{HS}$, is very low (see Fig. \ref{Fig:Comparison}, \new{panels E and F). As these panels show, GRNs show a range of values for the directionality of information flow that is wider than that shown by the Debian networks, and both systems share similarly-small fraction of information-unit pairs with a reciprocal dependence for a given network size; furthermore, the values for the different GRNs seem to follow the functional dependence between this fraction and network size shown by Debian.}

\begin{center}
\begin{figure}[hbtp]
\begin{centering}
\includegraphics[width=0.8\paperwidth]{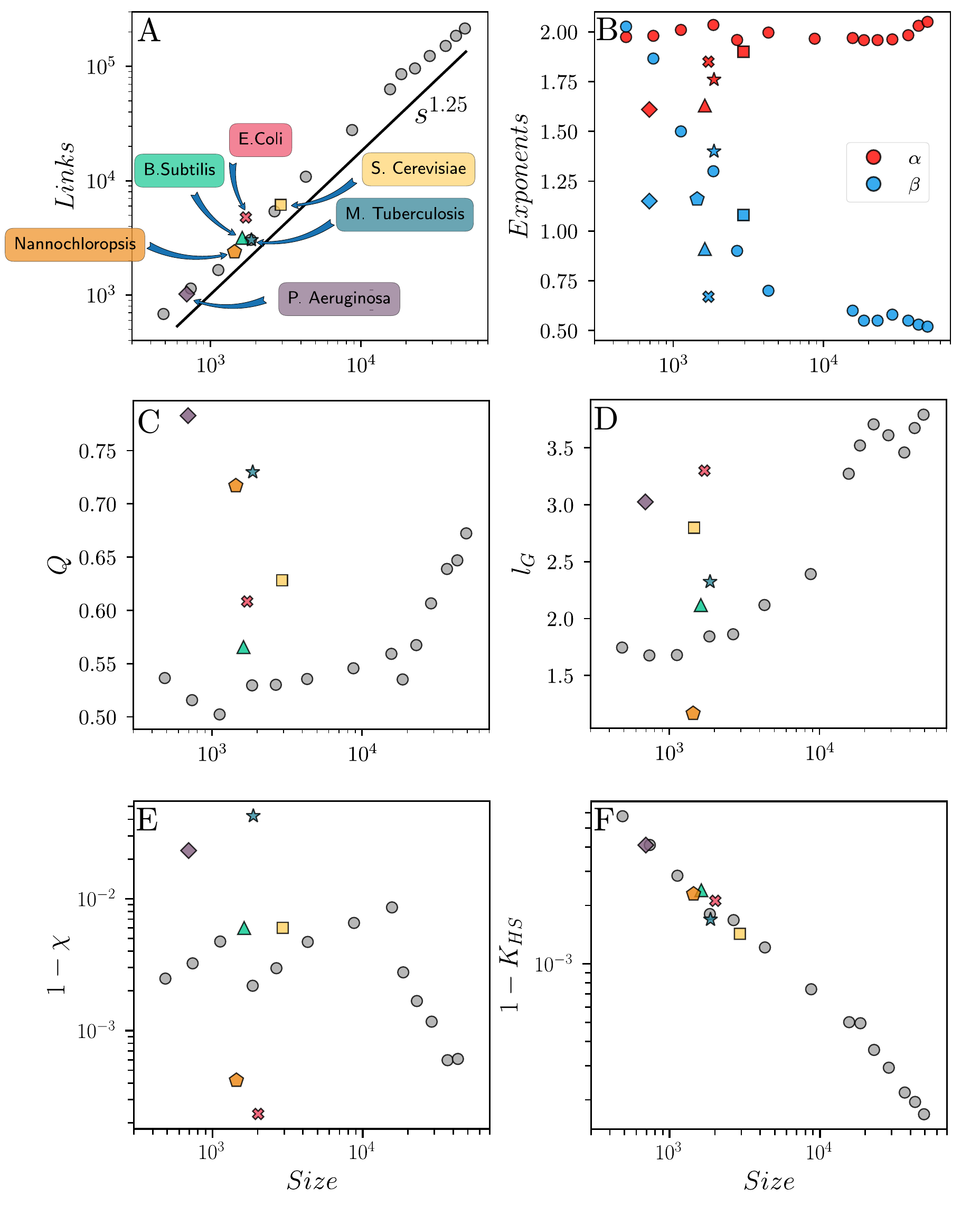}
\par\end{centering}
\caption{\new{Comparison between Debian and GRNs for different properties dependent on size (horizontal axis for all panels). A) Total number of links versus system size, with Debian networks represented by grey circles and GRNs by symbols. B) Exponents associated with the scale-free ($\alpha$) and stretched exponential ($\beta$) degree distributions mentioned in the previous sections. C) Newman modularity index (Q) for Debian networks (grey circles) and GRNs (symbols). D) Average path length versus system size for Debian networks (grey circles) and GRNs (symbols). E) One minus the flow index, $1-\chi$, for different Debian releases and GRNs as a function of the system size. F) One minus the Krackhardt hierarchy score, $1-K_{HS}$. \label{Fig:Comparison}}}
\end{figure}
\par\end{center}

Moreover, it is possible to quantify ``damage avalanches'' in GRNs. Gene knock-out experiments silence individual genes and monitor the cascade of genes that are affected by such a network-state change. The knock-out of a single gene is able to generate ``avalanches'' or  cascades of ``failures'' whose sizes follow a power-law  distribution with an exponent $\tau=\nicefrac{3}{2}$ (see Fig. S8).

We also explored how the information content of GRNs (defined as the amount of DNA per haploid nucleus in the genome, also called C-value), correlates with the number of genes in their genome for various organisms for which information was publicly available\cite{geer2009_ch6}. A positive correlation between the total number of genes and the genome size has been known for a long time to hold for many living organisms \cite{geer2009_ch6}, from prokaryotes to eukaryotes and plasmids, as well as viruses. \new{As illustrated in  Fig.\ref{FigCvalue}, our results confirm} this positive correlation. Moreover, the figure shows that viruses, plasmids, and prokaryotes share the same linear trend (i.e. the genome size of the organisms is not affected by organismal complexity). For eukaryotic cells, however, information content grows faster than the number of genes (i.e. the GRN contains much more DNA per gene than expected from the number of genes). This phenomenon has been called the ``C-paradox'' or, more recently, ``C-value enigma'' \cite{gregory2001}, and has been explained as caused by the existence of non-coding DNA in eukaryotes \cite{ohno1972}, i.e. a DNA segments that do not encode protein sequences but are believed to be important for control and regulation.

\begin{center}
\begin{figure}[hbtp]
\begin{centering}
\includegraphics[width=0.8\paperwidth]{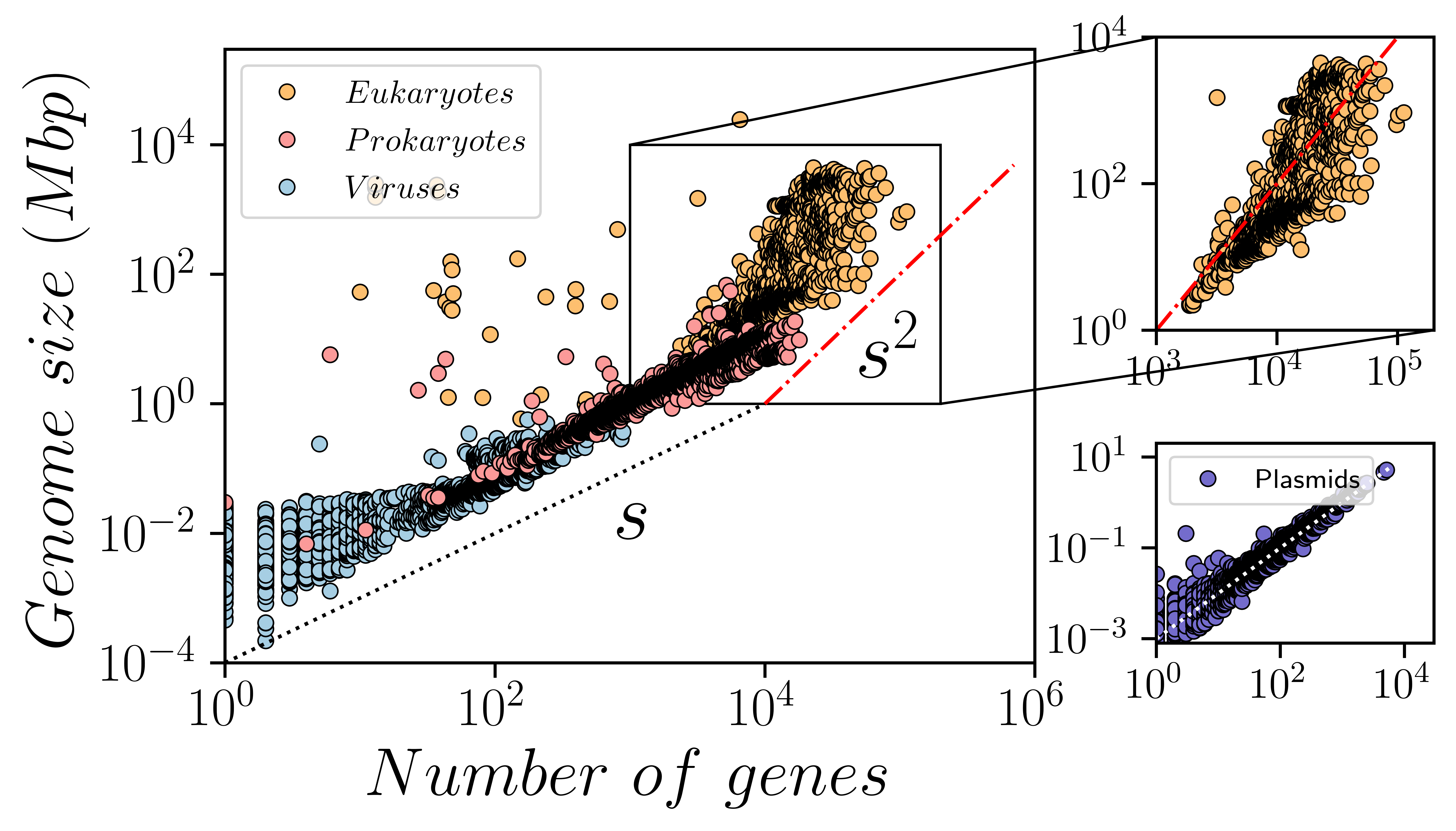}
\par\end{centering}
\caption{Genome size as function of the total number of genes (in log-log scale) for multiple microbial organisms (data from the NCBI database \cite{geer2009_ch6}).
  Observe the linear scaling associated with prokaryotes and viruses (red and blue points, respectively) as well as to plasmids (lower inset, violet points). On the other hand, data for eukaryotes depart from this linear relationship, becoming steeper to an approximately quadratic scaling (orange dots, enlarged in the upper inset). Lines are guides to the eye.
\label{FigCvalue}}
\end{figure}
\par\end{center}

\section{Discussion}

Tracking the evolution of Debian software networks offers an excellent opportunity to shed light and make conjectures on how information-processing biological networks could have possibly evolved.  Here, we explored the emergent architectural features of Debian networks achieved through their evolutionary dynamics. In particular, we showed that Debian networks are quite sparse, and show a scale-free structure for the out-degree distribution (representing the number of dependent packages) and a stretched-exponential in-degree distribution (representing the number of processing packages). In addition, the comparison of the releases with the ensemble of randomized networks allowed us to conclude that progressively more modular structures emerge from (and plays an important role in) the development of software networks, and that this is not a simple artifact of the particular degree distribution of the networks. The increase of the modular structure of the networks for the different releases is especially significant because it suggests an effective increase of their robustness, that is, in the ability to recover from malfunctioning or damaged packages\cite{Fortuna2011}. Moreover, the small average path length and its change over versions reveal that the typical distance for information transmission between any two packages remained short regardless of the growing network or their actual degree (i.e. small-world effects). 

In addition, the evolutionary trend for the three hierarchical categories identified here points to an increase in the re-use of code as new software was developed, which may have resulted from seeking a more efficient package use. As a consequence, Debian is becoming progressively more ``entangled'' \cite{Donetti}. The reuse of code (or software) to optimize the development of the operating system is, nonetheless, concerted with an effective and robust transfer of information, which is appreciable in the fact that the measures of hierarchy that refer to information flow show high directionality and low reciprocal dependencies. The potential for information flow improvement is compatible with the behavior observed for $l_G$, as well as the power law distribution observed for the cascade of failures originated from individual-package deletion. For the latter, an exponent $\tau=3/2$ (approximately what we observed for the last releases) is commonly interpreted as underlying critical un-biased branching process dynamics\cite{branching,ramo2006_ch6,MAM-rmp}, and is observed in many other contexts in biological systems \cite{MAM-rmp}. However, our results stem from structural effects, without requiring critical dynamics to explain them. Moreover, while the network has become more entangled the ``vulnerability'' of the system has decreased across releases. All together suggests that Debian has evolved towards a robust state in which code reuse is concomitant with an increase of the modularity of the network and a decrease of the risk of a collapse of the entire system\cite{Fortuna2011}. Although the increase in modularity in spite of the increase in the number of pass-through packages may seem surprising, it is understandable if this reutilization of resources seems to be done ``within moduli'' and thus preserving modular structure.

Interestingly, many of the structural properties discussed above seem to have changed their trend at a particular moment of the Debian evolutionary path. For example, the exponents of the in- and out-degree distributions, average in-degree, average path length, and modularity (of the swapped versions) all reach a stationary value around version $8$. This points to an evolutionary steady state (or, at least, stasis) for the network structure. At this particular version, we also observed an acceleration in the increase of modularity of the original networks and their total size, and a deceleration in the growth of the network (number of packages) and vulnerability of the network. Finally, also at this version, we observed high peaks of variability in different quantities (e.g. $Q$, $l_{G}$). These peaks indicate a large variability in the ensemble of randomized networks, which reveals some type of dramatic structural reorganization (transition) of the network ensemble. We were unable to track down any specific event/change in the development history of Debian that could explain solely all this phenomenology, although some important changes were introduced in the versions around and at the $8$th release. From a historical point of view, the 8th release saw dramatic design changes that meant to make the distribution more user-friendly, and more focused by reducing the broad spectrum of architectures covered by the operating system. This release introduced the $64$-bit architecture, and a more modular installation system (`debian-installer`) that included a tool to retrieve all dependencies for a specific package upon installation \cite{Hertzog2014}. At the same time, Ubuntu (a  distribution which aimed to improve the ease of the installation and use of Linux without compromising in robustness) was born, built on Debian's architecture and infrastructure. This distribution became very popular and its development led to a more professional/less hobbyist approach from developers, many of which created and maintained packages for both distributions. This feedback between Debian and Ubuntu (and change of mindset) may have also been important for the improvement in robustness that we report here. Thus, not only technical aspects but also environmental and sociological aspects have contributed to the evolution of Debian.

\subsection*{Analogies and differences between Debian and gene-regulatory networks}

On the basis of a reductionist position, the central dogma of molecular biology states that each isolated gene --i.e. the basic unit of heredity-- is transcribed into RNA. The latter, in turn, is usually translated into a protein, which can regulate the expression of other genes, usually inter-related with other proteins, thus conforming a complex network of regulatory interactions \cite{Dogma_ch6}. This entails causal relationships and the transfer of sequential information between such genes. Therefore, genetic expression is an essential biological process that can be represented in terms of GRNs \cite{buchanan2010_ch6}. Based on causal topological relations and the transfer of sequential information, we propose that the software networks of Debian constitute a useful analogy to understand some aspects of GRNs. Many of the features described here for Debian networks are similarly found in real GRNs.

The role of software packages in Debian networks is played by genes in GRNs, while the role of software dependencies is played by genetic regulatory dependencies or interactions. Both are typically sparse networks, with a similar structure reflected in their degree distributions: a scale-free structure for the out-degree distribution and a stretched-exponential in-degree distribution, a high level of modularity; and similar (small) values for the average path length indicating small-world effects. These similarities suggest the existence of some common underlying generative mechanisms for both types of networks, maybe focused on the common goal of information transmission. 


GRNs show a hierarchical structure, which has been hypothesized to confer an effective and robust way to transfer information and coordinate diverse processes \cite{Yan2010,wagner2007}, fostering the flow of information and providing effective responsiveness to external stimuli \cite{ravasz2002,babu2004}. Differently from Debian, however, the number of pass-through information units in GRNs is very reduced, showing a more pyramidal hierarchical structure. This is a key difference between the two systems. Although an advantage when it comes to the development of new individual packages, reuse of code should a priori decrease its robustness, as it facilitates the spread of damage. The suggestion that the reuse of code in Debian is kept within modules allows the operating system to grow with the advantages of both the efficiency of code reuse and modularity. Thus, both systems seem to have found ways to develop robustness, Debian aiming at a cost-efficient design principle \cite{Yan2010}. Nonetheless, both Debian and GRNs show a similar behavior for ``damage avalanches'', as the latter also show a power law distribution for avalanche size with an exponent close to $3/2$ (see above) \cite{ramo2006_ch6,MAM-rmp}.

Remarkably, the Debian software networks show a crossover from linear to \new{a much more accelerated} behavior (Fig. \ref{FigPack}) similar to that of the so-called ``C-value paradox'' for GRNs (Fig. \ref{FigCvalue}). For the latter, this change of slope is now known to result from the existence of non-coding DNA sequences. The functionality and role of such non-coding DNA, however, has not been fully elucidated and its utility or futility are still under discussion, even if it seems clear that it is important for gene regulation \cite{carey2015,mercer2009}. However, it is important to stress that some studies have revealed the tendency of the genomic size to increase evolutionarily because mutation sometimes duplicates parts of the genome (i.e. by means of nonadaptive processes without, in principle, any functional meaning). Some of these duplicated genes were removed by natural selection (purging non-beneficial sequences) while some others resulted in novel substrates for the evolution of further complexity and functionalities, possibly including a role in regulatory functions \cite{conery}. Interestingly, looking into the structure of individual Debian packages, we found extra pieces of information that are present in each package beyond the essential information required to work. In particular, handbooks or extra control information encapsulated in particular control files, which contain meta-information such as the needed dependencies, as well as mechanisms designed to avoid errors during unpacking and installation processes. Such information ensures a proper transmission of information during the installation process, i.e. ensures that packages are really functional in the whole network\cite{krafft2005_ch6}. This conforms for software the equivalent of non-coding pieces of DNA in GRNs. In this analogy, at least one of the roles of non-coding pieces is the monitoring and minimization of gene disruption during the transcriptional process. We hypothesize that control files and extra information comprise and ensure the resilience of the network, minimizing risks in the transmission of the information of the system. The increased growth of the amount of information in the highly-complex networks of the last releases points to an increasing relevance of control files as an effective strategy in order to minimize the disruption during information processing. This could allow for an emergent optimized structure for the flow of information throughout the network. Although all these hypotheses can help understand the function and evolution of non-coding DNA, further work is needed to clarify the role and changes of control files over releases, as well as the evolution of the internal structure of software packages.

\section{Conclusions}

As shown here, in many aspects Debian networks are able to recreate some of the emergent properties observed in real gene regulatory processes \cite{Fortuna2007,Fortuna2011,Yan2010}. In both synthetic and real biological networks, common solutions emerge to the general problem of designing circuitry that optimizes storage, information processing (of both internal and external stimuli), and robustness (see Table I). Debian has evolved towards a stable scale-free out-degree distribution that is also shown by GRNs. In Debian, this property emerges from a network that grows by adding packages that are not necessarily needed by others but that require using key existing ones. For GRNs, this would mean that the genome grows by adding new genes that are controlled by key, existing genes, opening thus new regulatory pathways. In fact, empirical evidence from analysis of E. coli and yeast GRNs reveals that duplication of target genes has contributed far more to the growth of their GRNs than duplication of transcription factors (see e.g. \cite{babu2}). The exponential in-degree distribution in Debian resulted from a self-regulated limitation as to how much of the existing network is reutilized by new packages, i.e. there is a finite, well-defined mean number of dependencies per package. For GRNs, this self-regulation is convenient as it reduces the vulnerability of the network. The evolution of Debian has increased its robustness through a reduced average path length and increased modularity, which could be a key reason why current GRNs, also showing low $l_G$ and high $Q$, have been selected for. The two types of networks, however, have found different strategies to increase their robustness. Whereas the growth of the Debian network has required a within-moduli reuse of packages, GRNs have a pyramidal hierarchy that could have resulted from adding new genes that introduce functional redundancy (as opposed to re-using existing ones), which considerably reduces the vulnerability of the network.

The study of similarities and differences between Debian and GRNs has helped here understand better the evolution of the latter. Conversely, it might be of interest to consider borrowing and implementing some features and strategies from real gene regulatory networks for the development of operating systems. In future work, we plan to develop a mathematical model of growing networks aimed at accounting for these common architectural features between software and gene regulatory networks, which can help identify the origin and nature of the divergences between them. Moreover, we plan to extend this study to other Linux distributions and other software ecosystems (e.g. Python Package Index or R-CRAN), with the aim to unveil key commonalities and differences that help us understad the different evolutionary solutions to common developmental challenges. Similarly to the present study, such an analysis can contribute to our understanding of the evolutionary processes that current self-organized biological networks have undergone or will undergo in the future.

\section{Methods}

\paragraph{Debian networks.}

The Debian GNU/Linux networks are composed by software packages (units of software), which can depend on other packages (the so-called dependencies) and/or exhibit conflicts with other packages. A dependence implies a direct link (from package $A$ to package $B$) that means that package $B$ should be installed first to be able to install $A$ (and, consequently, information flows from $B$ to $A$). For conflicts, the link means that package $A$ cannot be installed if package $B$ is present (i.e. installed) in the system. 

Conflicts are mostly used to avoid duplicities that can ensue from the different options for the same requirement or functionality (grouped in the so-called ``virtual packages''), as well as to select between previous versions of different packages, as explained in the Debian Policy Manual \cite{DPolicy}. Conflicts between virtual facilities can be solved by considering particular choices among the various possibilities. Thus, in order to build Debian potential networks excluding duplication of the so-called $virtual$ packages, here we selected randomly a particular choice of each possible real package providing such functionality, and obtained averages over choices (specifically, over $10^{4}$ realizations for every distribution), hence exploring the whole ensemble of networks for each release. As a result of solving such conflicts, these Debian potential networks indicate the requirements (or dependency network) between the different packages.

\paragraph{Statistical analysis.}

We performed curve fits for the cumulative degree distributions for dependencies, employing a Levenberg--Marquardt algorithm\cite{more1978}. For cases compatibles with power laws, we used cumulative distributions for this curve-fitting; in other words, if the expectation is a degree distribution of the form $P(k)\sim k^{-\alpha}$, we studied instead its cumulative version, ($cP(k)=\operatorname{P}(K\leq k)=\int_{k_{min}}^{K}P(k)dk \sim cP(k)\sim K^{-\alpha+1}$). On the other hand, when the expectation was a stretched exponential (.e.g in-degree distribution), that is $P(k)\sim exp(-(\frac{k}{\tau})^\beta)$, we instead curve-fitted its transformation $log P(k)\sim-(\frac{k}{\tau})^\beta$.

In order to compare modularity and another properties across releases (see below), we also computed the Z-Score with respect to random networks, with random realizations respecting the degree sequence of the network (i.e. the number of in- and out-neighbors for each package). The Z-score of an observable in the network is defined as the difference between the observable and the mean, weighed by standard deviation of the randomization: $z=\frac{x-\mu}{\sigma}$.

\paragraph{Network swapping.}
Given the original degree sequence of a Debian network ($D$), the randomization of the Debian potential networks is performed by making modifications on a copy ($D'$). Such modifications are no more than a swapping process that maintains both the in-degree and the out-degree connectivity. The swap of the links is performed as follows:
\begin{enumerate}
\item Select randomly a package $A$ and one of its dependencies $a$.
\item Select  randomly a different package $B$ and one of its dependencies $b$ with the  conditions:
      (i) $A$ cannot depend on $b$ and (ii) $B$ cannot depend on $a$.
\item Swap both links, i.e. $A$ depends on $b$ and $B$ depends on $a$.
\item Iterate this process a large number of times until the  network becomes fully randomized, i.e. until the observable in the randomized network reaches a stationary state.
\end{enumerate}

Once this process has been completed, the result is a randomized version of the focal Debian network/release that can be used to measure all the desired structural features, with the confidence that the networks are comparable. Here, we used an ensemble of over $10^{4}$ realizations for every release. 

\paragraph{Network structural features.}

To measure of the modularity index, $Q$ in the different releases of Debian we used a heuristic method, the Louvain method\cite{Blondel2008} on directed networks \cite{Dugue2015}, based on modularity optimization. The modularity index is defined as \cite{newman2003_review}:

\begin{equation}
Q=\frac{1}{m}\underset{i,j}{\sum}\left[A_{ij}-\frac{d_{i}^{in}d_{j}^{out}}{m}\right]\delta\left(c_{i},c_{j}\right)
\end{equation}

where $A_{ij}$ is the adjacency matrix, $m$ the number of edges, $d_{i}^{in}$ (respect. $d_{i}^{out}$) stands for the in-degree (out-degree) of the node $i$ and $c_i$ the community to which vertex $i$ belongs. The Louvain method provided in this case an excellent accuracy due to the relatively large size of the last Debian releases.

On the other hand, the average path length on the graph ($l_{G}$, i.e. the number of steps along the shortest paths for all possible pairs of network nodes), is defined as:

\begin{equation}
{\displaystyle l_{G}=\frac{1}{m\cdot(m-1)}\cdot\sum_{i\neq j}d(n_{i},n_{j})} 
\end{equation}
where $m$ is the number of links and $d(n_{i},n_{j})$ is the shortest distance between nodes $i$ and $j$. Here, we measured $l_{G}$ using the \textit{igraph R} package\cite{Csardi2006}. 

To quantify hierarchy, we used three different indicators. First, we measured the number of sinks, sources, and pass-through packages, which provides an indication of hierarchical level and re-use of packages in the network. In addition, we measured an index that focuses on the directionality of the information flow in the network \cite{VirIndex}, the so-called flow index $\chi$, which measures the fraction of links pointing from lower to higher hierarchical levels, i.e. aligned with the inherent directionality. Computing $\chi$ requires the determination {\it a priori} of a hierarchical ordering of all nodes in the network for which one can use the algorithm developed by Dominguez-Garcia et al. in \cite{VirIndex}. The index takes the value $\chi = 1$ in the limit of perfect feedforwardness (perfect directionality or hierarchical organization), and $\chi=1/2$ in the absence of a well-defined directionality (e.g. for a random directed network). For convenience when using a semi-logarithmic plot, we represented in our figures $1-\chi$ instead. Lastly, we measured a version of the classical hierarchy index that quantifies the level of mutual dependencies. The Krackhardt hierarchy score, $K_{HS}$ is defined as the fraction of (unordered) node pairs $(i,j)$ such that node $i$ is reachable from node $j$ but node $j$ is {\it not} reachable from node $i$ or viceversa \cite{Krackhardt}: 
\begin{equation}
K_{HS}=1-\frac{V}{N(N-1)}
\end{equation}
where $V$ is the number of symmetrically linked pairs and $N(N-1)$ is the total number of pairs in the network. Here, we instead represented $1-K_{HS}$, which indicates the fraction of pairs that show mutual dependencies.

\section*{Acknowledgments}
We thank E. Estrada, P. Moretti and J. M. Mart\'in for very useful comments. We are also grateful to two anonymous reviewers, whose comments and suggestions have greatly improved this manuscript. We acknowledge the Spanish-MINECO grant FIS2017-84256-P (FEDER funds) for financial support, as well as the Consejer\'ia de Conocimiento, Investigaci\'on y Universidad, Junta de Andaluc\'ia and European Regional Development Fund (ERDF), ref. SOMM17/6105/UGR. 
\section*{Author Contributions}
J.A.B. and M.A.M. conceived the project, P.V. performed the numerical simulations and prepared the figures. P.V., J.A.B. and M.A.M. wrote
and reviewed the manuscript.


\bibliographystyle{naturemag}
\def\url#1{}



\section*{Supplementary material}

# Supplementary Information for "Evolution in the Debian GNU/Linux software network: analogies and differences with genetic regulatory networks"

**Pablo Villegas**[1,2]**, Miguel A. Muñoz**[1]**, and Juan A. Bonachela**[3,4,*]

[1]Departamento de Electromagnetismo y Física de la Materia e Instituto Carlos I de Física Teórica y Computacional. Universidad de Granada, E-18071 Granada, Spain
[2]Istituto dei Sistemi Complessi, CNR, via dei Taurini 19, 00185 Rome, Italy
[3]Marine Population Modeling Group, Department of Mathematics and Statistics, University of Strathclyde, Glasgow, G1 1XH, Scotland, UK
[4]Department of Ecology, Evolution, and Natural Resources, Rutgers University, New Brunswick, New Jersey, United States of America
[*]To whom correspondence should be addressed: juan.bonachela@rutgers.edu

## SI-1: Properties of the Debian networks

The growth of the total number of packages in the Debian/GNU Linux operating system across its first 14 releases is summarized in Figure S1. The number of packages increased exponentially over time showing two different trends, with the 8th release as the transition point.

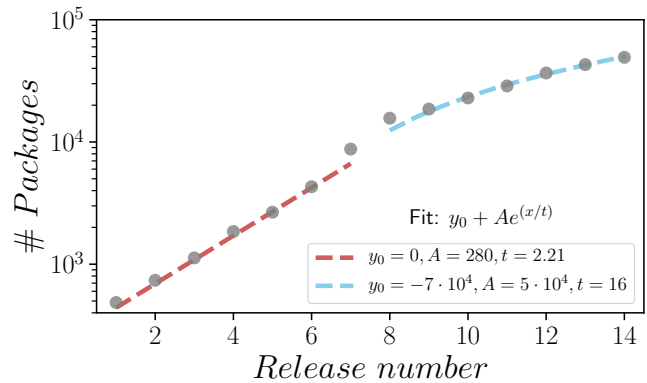

**Figure S1.** Total number of packages versus release number in Debian GNU/Linux networks. Two different exponential growths can be appreciated: One before release number 8 (red dashed line), and another one thereon (blue dashed line).

## SI-2: Probability distributions of Debian networks

We also monitored the out-degree distribution by measuring the cumulative probability distribution for a package to be a requirement for $k_{out}$ packages for each distribution. As shown in Fig.S2 the cumulative version of the out-degree distribution follows a power law, $P(k_{out}) \sim k_{out}^{-\alpha+1}$ with characteristic exponent $\alpha$. As can be seen, the power law of the out-degree distribution shows an exponent very close to $\alpha = 2$ for all releases (red points represents the probability distributions and black lines are fits with an exponent is indicated for each case).

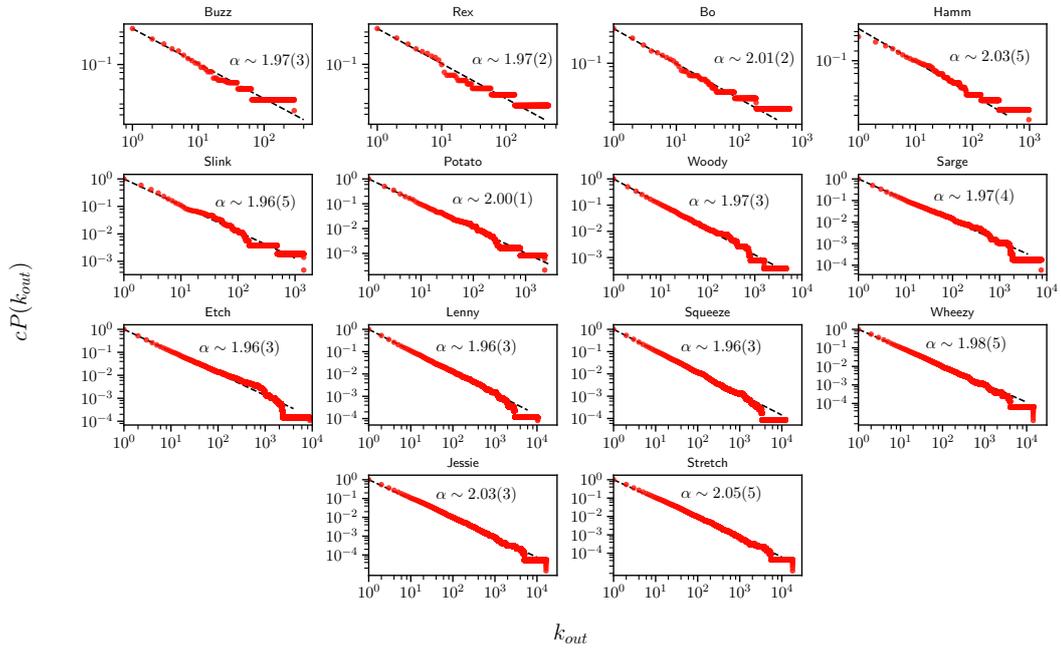

**Figure S2.** Cumulative degree distribution for outgoing dependencies for all releases (red points) and associated fits (black dashed lines).

On the other hand, the probability for a package to depend on $k_{in}$ packages follows a stretched exponential, $P(k_{in}) \sim \exp(-(\frac{k_{in}}{\tau})^{\beta})$ with characteristic exponent $\beta$. The exponent of the stretched exponential for the in-degree distributions decays from $\beta = 2$ (normal distribution) to an approximately-stationary $\beta \approx 0.5$ after the 8th release (blue points).

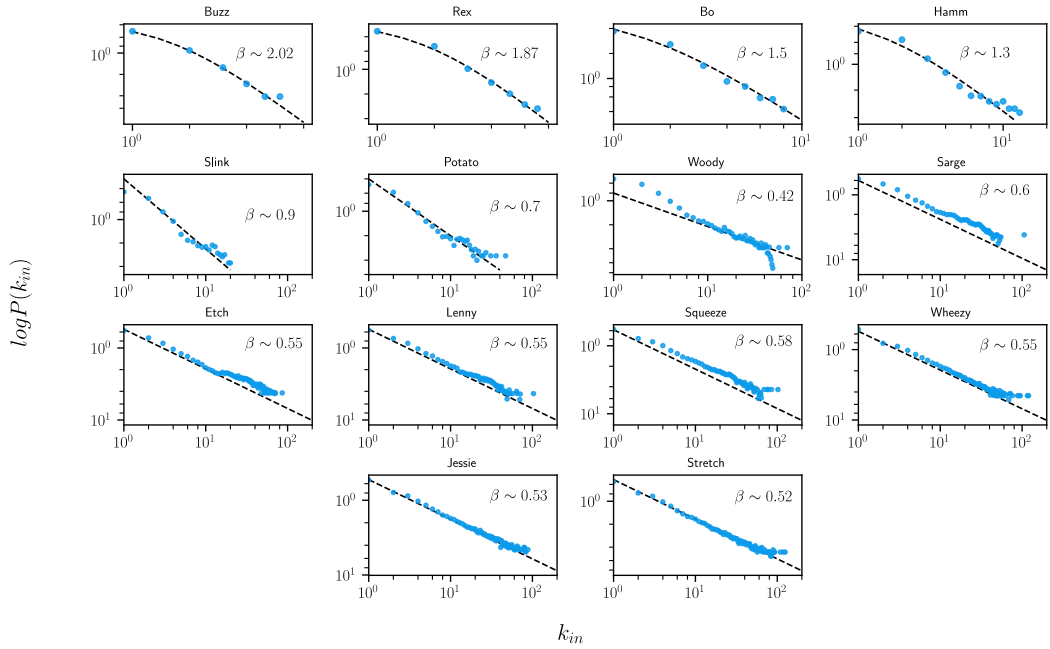

**Figure S3.** Logarithm of the degree distribution for incoming dependences for all releases (blue points, note the inversion of the vertical axis), showing the change from a half-normal distribution decay ($\beta = 2$) for the first version to a stretched exponential distribution ($\beta = 0.5$) for the last one. Dashed black lines are fits of the form $logP(k_{in}) \sim k^{\beta} + h$.

## SI-3: Measures in Debian networks

We also computed the Z-Score with respect to random networks respecting the degree sequence of the network (i.e. the number of in and out neighbors for each package). The Z-score of an observable for each network is defined as the difference between the observable and its mean across randomizations, taking into account its variability (i.e. normalizing by the standard deviation). Mathematically, $z = \frac{x-\mu}{\sigma}$. Figure S4 shows the Z-Score values for the modularity index (Z-Score Q) and the average path length (Z-score $l_G$). Both values are far from $Z \sim 0$, indicating deviations on the order of $(500-800)\sigma$ for the modularity index in the last distributions (i.e. the networks show a modularity index much larger than expected in the random case), and on the order of $(10-30)\sigma$ for the average path length in the last distributions (i.e. the networks show an average path length much shorter than expected in the random case). In addition, Fig. S5 shows the growth of the average path length as a function of the logarithm of the total number of packages. For Debian networks, and as expected for small-world networks, $l_G$ exhibits a growth proportional to the logarithm of the size of the network, i.e. $log(N_{pack})$. As stated in the main text, there exists a shift in behavior –around the 8th distribution– where all the major changes take place.

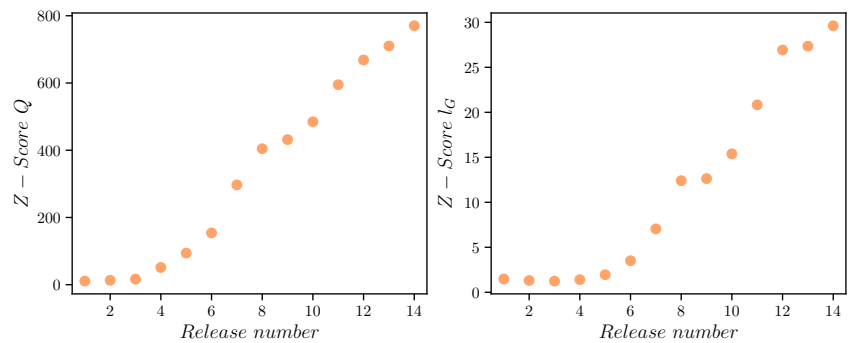

**Figure S4.** Left: Z-score value for the modularity index versus release number; modularity shows a sustained increase over time with respect to the expected value of the "swapped" network. Right: Z-score value for the average path length versus release number; the average path length remains much lower than expected as the system size increases in relation to the "swapped" networks.

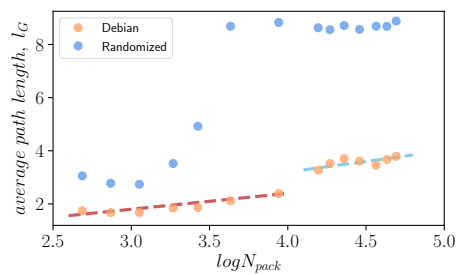

**Figure S5.** Average path length ($l_G$) for the original Debian networks (orange points) and the "swapped" networks (blue points) as a function of the log-number of packages. Dashed lines (red and blue) stand for different growths proportional to $log(N_{pack})$, before and after the 8th distribution.

## SI-4: Probability distributions of gene regulatory networks

We also measured the cumulative probability distribution for regulated genes in diverse gene regulatory networks (GRNs) for which data were publicly available (see main text). As shown in Fig.S6, the cumulative version of the out-degree distribution follows a power law $P(k_{out}) \sim k_{out}^{-\alpha+1}$ with characteristic exponent $\alpha$, at least for a few decades. The associated exponents are in the range $\alpha \in (1.5 - 2)$.

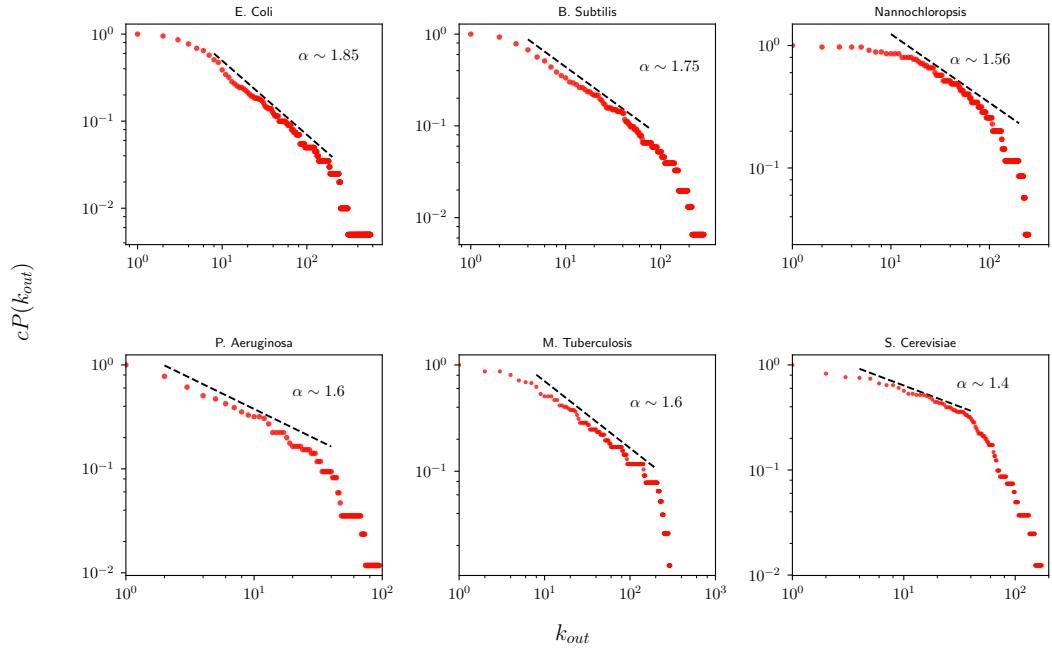

**Figure S6.** Cumulative degree distribution for regulated (i.e. outgoing) genes for different GRNs (red points), with guides to the eye for the power-law fit (black dashed lines).

On the other hand, the probability for a gene to be regulated by $k_{in}$ genes follows a stretched exponential, $P(k_{in}) \sim \exp(-(\frac{k_{in}}{\tau})^{\beta})$ with characteristic exponent $\beta$ (see Fig. S7). The exponent of the stretched exponential follows $\beta = 1$ (i.e. exponential distribution) except for *E.Coli*, for which $\beta \approx 0.67$.

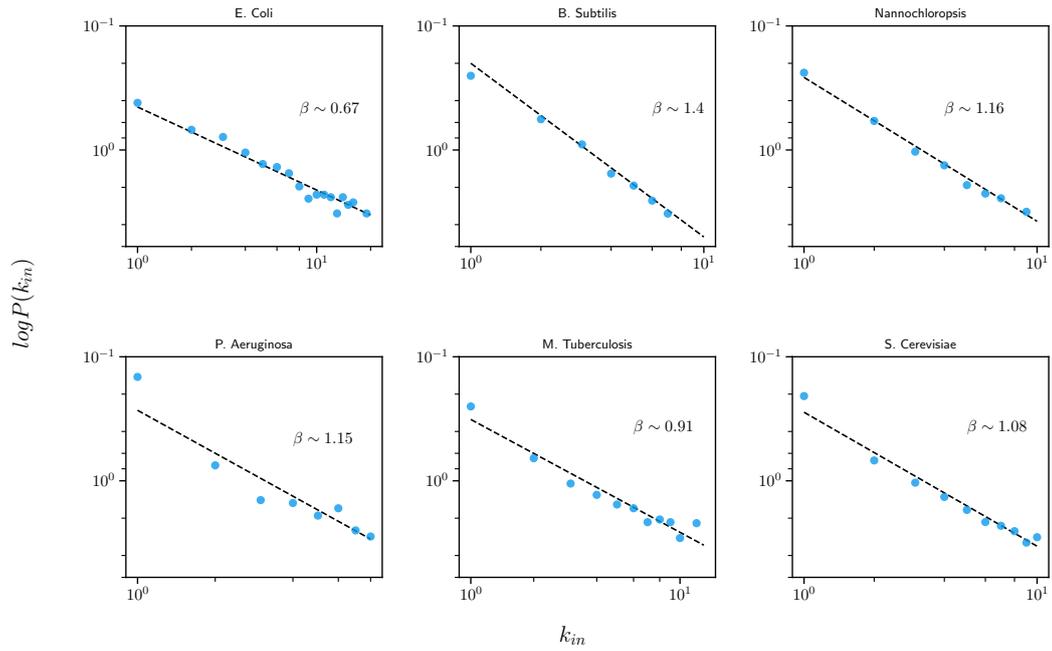

**Figure S7.** Cumulative degree distribution for regulatory (i.e. incoming) genes in different gene regulatory networks (blue points, note the inverted vertical axis), with guides to the eye for the exponential fit (black dashed lines).

## SI-5: Further comparison between GRNs and Debian networks

Gene knock-out experiments silencing individual genes allow for a measure of the "vulnerability" (see main text) of GRNs by monitoring the cascade of metabolic changes that are affected by such a network-state change. As shown in Figure S8, the distribution of the size of such cascade follows a power-law distribution with an exponent $\tau \sim \text{3/2}$, similar to that measured for the Debian networks. See main text for further details.

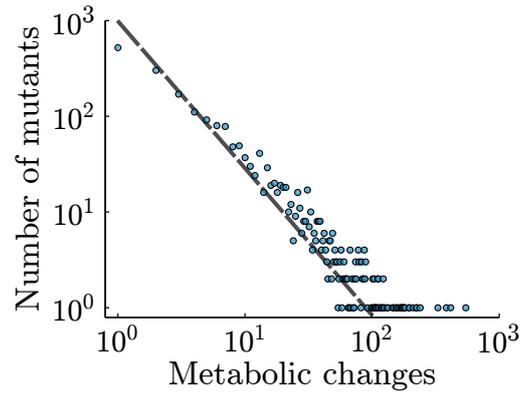

**Figure S8.** Distribution of metabolic changes in gene knockout mutants for E.Coli (blue points). Adapted from Fuhrer et al[1]. The dashed line is a power law with exponent $\tau = \frac{3}{2}$, as guide to the eye.



\end{document}